\documentstyle[lanl,pslatex,rawfonts]{article}  
\begin{document}
% -----------------------------------------------------------------------%| 
% Template TeX file for                                                  %| 
%                                                                        %| 
%       17th Winter Workshop on Nuclear Dynamics                         %|
%                                                                        %|
%       Park City, Utah, USA, March 10-17, 2001                          %|
%                                                                        %|
% Please, keep this header when composing your own TeX source.           %|
%                                                                        %| 
\frompage{000} \topage{000}                                              %| 
%________________________________________________________________________%|  
\def\rightmark{$Au + Au \rightarrow Au + Au + \rho^0$} 
\def\leftmark{Klein et al.}
 
\title{Observation of $Au + Au \rightarrow Au + Au + \rho^0$ \\ 
and $Au + Au \rightarrow Au^* + Au^* + \rho^0$ with STAR}

\authors{
{Spencer Klein$^1$ for the STAR collaboration$^a$}\\[2.812mm]
{\normalsize
\hspace*{-8pt}$^1$ Lawrence Berkeley National Laboratory \\ 
Berkeley, CA, 94720, USA\\
SRKlein@lbl.gov\\[0.2ex] 
}}

\abstract{We report the first observation of the reactions $Au + Au
\rightarrow Au + Au + \rho^0$ and $Au + Au \rightarrow Au^* + Au^* +
\rho^0$ with the STAR detector.  The $\rho$ are produced at small
perpendicular momentum, as expected if they couple coherently to both
nuclei.  We discuss models of vector meson production and the
correlation with nuclear breakup, and present a fundamental test of
quantum mechanics that is possible with the system.}



\maketitle

\section{Introduction}\label{intro}

With their large charge ($Z$), relativistic heavy ions can interact
electromagnetically, even at impact parameters $b$ much larger than
twice the nuclear radius $R_A$.  In these ultra-peripheral collisions,
the two nuclei act as sources of fields.  The electromagnetic fields
have a long range, so two-photon and photon-Pomeron interactions have
large cross sections.  Exclusive Pomeron-Pomeron reactions have a
small cross section because of the short Pomeron range.  The
amplitudes for the photon or Pomeron emission from each nucleon add.
The momentum transfer is small enough that the amplitudes all have the
same phase, so the coupling is coherent over the entire nucleus.  The
cross section for two-photon interactions goes as $Z^4$ ($4\times
10^7$ with gold ions). For photon-Pomeron interactions, $\sigma\sim
Z^2A^2$ for 'heavy' states like $J/\psi$ and $\sigma\sim Z^2A^{4/3}$
for lighter mesons.  This scaling leads to large cross sections.  The
coherence constrains the fields to have perpendicular momentum $p_T <
\hbar/R_A$ and longitudinal momentum $p_{||} < \gamma\ \hbar/R_A$.
Final states can have masses up to $2\gamma\ \hbar/R_A$ (6 GeV at
RHIC). The final state $p_T$ is less than $\sim 2\hbar/R_A$.

\begin{figure}[htb]
\begin{center}
\leavevmode
\epsfclipon
\epsfysize=4.5  cm
\epsfbox{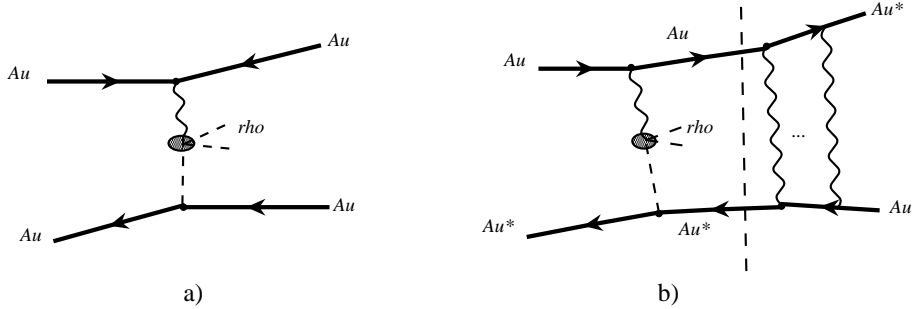}
\end{center}
\vskip -.25 in
\caption[]{The Feynman diagrams for (a) $\rho^0$ production, and (b)
$\rho^0$ production with nuclear excitation.  Although other diagrams
can lead to $\rho^0$ production with nuclear excitation, (b) is
believed to be dominant. The dashed line shows the factorization
following Eq. (\ref{eq:factor}).  The number of photon lines in (b) is
not fixed.  Two photons are shown to demonstrate how the excitations
of the two nuclei are independent.}
\label{fig:feynman}
\end{figure}

Photon-Pomeron or photon-meson interactions produce vector mesons, as
in Fig. \ref{fig:feynman}(a).  The Pomeron is a colorless object with
the same quantum numbers as the vacuum.  Photon-Pomeron interactions
can be described in terms of elastic scattering.  A photon from the
electromagnetic field of one nucleus fluctuates to a $q\overline q$
state.  This state then elastically scatters from the other nucleus,
emerging as a vector meson.  Vector meson dominance allows us to
calculate the rate for this process by treating light $q\overline q$
states directly as vector mesons. At RHIC, $\rho^0$, $\omega$, $\phi$
and $J/\psi$ are copiously produced, and the $\psi'$ should be
observable.  The upsilon family may be observable with lighter ion
beams.

Two-photon interactions include $e^+e^-$ pair production, single meson
production, and meson pair production.  Since the coupling $Z\alpha$
($0.6$ for gold) is large, $e^+e^-$ pair production is an important
probe of quantum electrodynamics in strong fields.  Coupling to
2-photons is sensitive to a mesons internal charge (quark or gluon
content).  These physics topics are reviewed
elsewhere\cite{rev1}\cite{capri}.

These reactions can be studied experimentally by selecting events with
low multiplicities and small total $p_T$.  We present a study of
$\rho^0$ production with and without nuclear excitation, based on data
taken with the STAR (Solenoidal Tracker at RHIC) detector at RHIC at a
center of mass energy of $\sqrt{S_{NN}} = 130$\ GeV per nucleon pair.
STAR is a collaboration of about 320 physicists from 31 institutions
in 8 countries.

\section{Expected Rates}

The cross section to produce a meson $V$ is the convolution of the
photon spectrum from one nucleus with $\sigma(\gamma A\rightarrow
VA)$\cite{vmprod}. The photon flux from one nucleus is given by the
Weizs\"acker-Williams virtual photon approach; the small photon
virtuality is negligible here.  For a photon-Pomeron or two-photon
interaction to be visible, there cannot be an accompanying hadronic
interaction.  This requirement is similar (but not identical) to
requiring $b > 2R_A$; it reduces the photon flux by roughly a factor
of 2.

The photon-nucleus cross section can be determined by a Glauber
calculation that uses the $\sigma(\gamma p \rightarrow Vp)$ as input.
The rate depends on how vector mesons interact with nucleons.  The
RHIC vector meson production rates can be used to determine the vector
meson-nucleon cross section.  The cross sections are very large, 380
mb for $\rho^0$ production with gold ions at $\sqrt{S_{NN}} = 130$\
GeV.  This is about 5\% of the hadronic cross section.  At
design energy and luminosity, RHIC is expected to produce 120 $\rho^0$
each second.  The production rates are high enough that rare meson
decays should be accessible, and excited vector mesons such as the
$\rho^*$, $\omega^*$ and $\phi^*$ can be studied.

Photons can also fluctuate to virtual $\pi^+\pi^-$ pairs.  One of
the $\pi$ can interact with the target nucleus, and the pair can
become real.  The amplitude for $\gamma\rightarrow \pi^+\pi^-$ is
independent of the pair invariant mass $m_{\pi\pi}$.  Direct
$\pi^+\pi^-$ production interferes with $\rho^0$ production, so their
amplitudes add\cite{soding}; the cross section is
\begin{equation}
{d\sigma\over dM_{\pi\pi}} = \bigg| 
{A\sqrt{M_{\pi\pi}M_\rho \Gamma_\rho} \over M_{\pi\pi}^2 - M_\rho^2
+iM_\rho\Gamma_\rho} + B \bigg|^2
\label{eq:rhosigma}
\end{equation}
where $A$ and $B$ are the amplitudes for $\rho$ and $\pi\pi$
production.  The $\rho$ width $\Gamma_\rho$ must be corrected for the
increasing phase space as $m_{\pi\pi}$ increases.  Several methods
have been proposed; we take $\Gamma_\rho = \Gamma_0 (p^*/p_0^*)^3
M_\rho/M_{\pi\pi}$, where $p^*$ is the pion momentum in the $\pi\pi$
rest frame, and $p_0^*=p^*$ when $M_{\pi\pi}=M_\rho$\cite{jackson}
and $M_\rho = 768$ MeV/c$^2$, $\Gamma_\rho = 151$ MeV/c$^2$\cite{pdg}.
The $\pi^+\pi^-$ spectrum also includes a small component from
$\omega\rightarrow\pi^+\pi^-$ which is neglected here.

As Eq. (\ref{eq:rhosigma}) shows, the $\rho$ and $\pi\pi$ interference
is constructive for $m_{ \pi\pi} < m_\rho$.  Around $m_{\pi\pi} =
m_\rho$, the $\rho$ component shifts phase by $180^o$, and at higher
masses, the interference is destructive.  This skews the overall $m_{
\pi\pi}$ distribution.

\subsection{Vector Meson production with Nuclear Excitation}

Vector meson production can be accompanied by nuclear excitation, as
in Fig. \ref{fig:feynman}(b).  For this, diagrams like
Fig. \ref{fig:feynman}(b) are expected to be dominant. One or more
additional exchanged photons can excite one or both nuclei to a giant
dipole resonance.  Higher collective excitations or nuclear breakup
are also possible.  However, photons directly involved in two-photon
or photon-Pomeron interactions are unlikely to also excite the
emitting nucleus\cite{hencken}.  Pomeron emission (elastic scattering)
is less likely to cause collective excitation than photon emission
because Pomerons couple identically to protons and neutrons.
Neglecting other production diagrams, the production probability
factorizes and
\begin{equation}
\sigma = \int d^2 b P_\rho(b) P_{2GDR}(b) [1-P_{HAD}(b)]
\label{eq:factor}
\end{equation}
where $P_\rho(b)$ is the $b$ dependent probability of production a
$\rho$, around 0.5\% for $b=2R_A$.  The $b$ dependence comes entirely
from the photon flux.  

The single nucleus excitation probability is $P_{GDR}(b) =
1-\exp{(-S/b^2)}$\cite{rev1}, where $S$ includes terms for different
types of photo-disintegration.  GDR excitation is the largest, but
higher excitations and other photonuclear reactions also contribute.
For gold at RHIC, $S\sim 150$\ fm$^2$.  The probability of exciting
both nuclei, $P_{2GDR}(b) \sim P_{GDR}(b^2)$, with
$P_{2GDR}(b=2R_A)\sim 30\%$\cite{baltz}.  These relationships assume
that the excitation of the two nuclei occurs independently.

$P_{HAD}$ is the hadronic interaction probability, which can be
approximated as 1 for $b<2R_A$; 0 otherwise.  The cross section for
vector meson production with nuclear excitation is expected to be more
than an order of magnitude smaller than for exclusive vector meson
production.

\subsection{Interference and the Vector Meson $p_T$ Spectrum}

Photons have a long range, and Pomerons (elastic scattering) have a
short range, so vector meson production must occur inside or near
(within 1 fermi) one of the nuclei.  Either nucleus can serve as the
emitter or scatterer; the two possibilities are indistinguishable, so
their amplitudes add and the two sources interfere.  At mid-rapidity
($y=0$), the contributions from the two sources are equal.

\begin{figure}[htb]
\begin{center}
\leavevmode
\epsfclipon
\epsfysize= 5 cm
\epsfbox{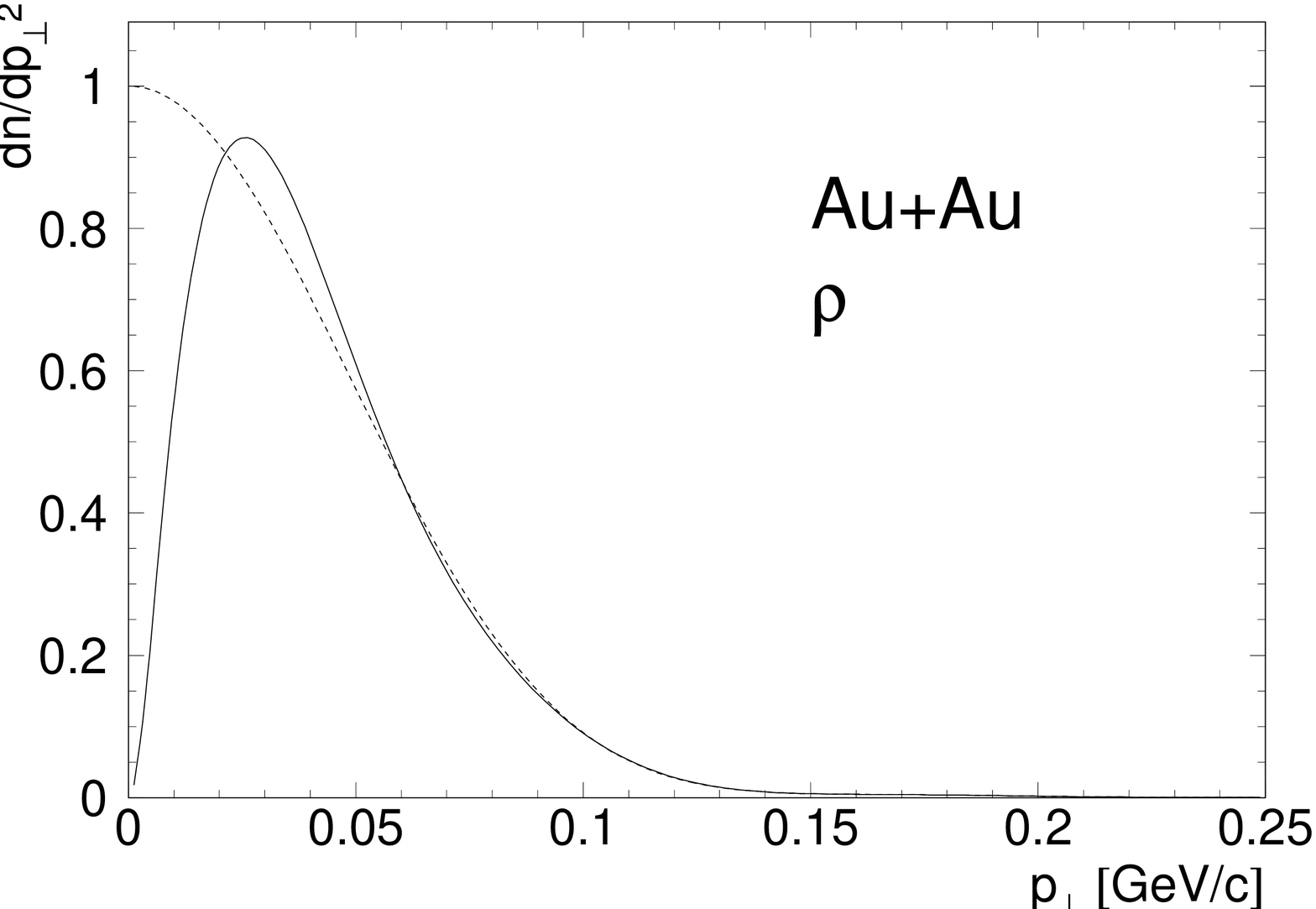}
\end{center}
\vskip -.2 in
\caption[]{Expected $\rho^0$ $p_T$ spectrum for gold on gold
collisions at 200 GeV per nucleon, with (solid line) and without
(dotted line) interference.}
\label{fig:interfere}
\end{figure}

Vector mesons
have negative parity, so the cross section at $y=0$ is\cite{vminter}
\begin{equation}
\sigma(p_T,y=0,b)= 2A^2(p_T,y=0,b) (1-\cos[\vec{p}\cdot
\vec{b}]).
\label{eq:sigma}
\end{equation}
where $A$ is the production amplitude, $\vec{b}$ the impact parameter
vector, and $\vec{p}$ the vector meson momentum.  For a given $b$,
$\sigma$ oscillates with period $\Delta p_T = \hbar/b$.  For
$\vec{p}\cdot\vec{b}\ll 1$, the interference is destructive, and
$\sigma$ is small.  Of course, $b$ is not observable in heavy ion
collisions, so Eq. (\ref{eq:sigma}) must be integrated over all $b$.
When this is done, the oscillations wash out except for $p_T < \langle
b \rangle$, where the cross section drops sharply.
Fig. \ref{fig:interfere} shows the expected cross section as a
function of $p_T$.

This interference is of special interest because vector mesons decay
quickly, before travelling the distance $b$ required so their wave
functions can overlap. The decay distance $\gamma\beta c\tau$ is less
than 1 fermi, far less than the $\langle b\rangle \sim 40$ fermi at
RHIC (and $\langle b\rangle \sim 300$ fermi at the LHC).  So, the
vector mesons cannot interfere.  However, their decay products can
interfere, as long as the system wave function retains information on
all possible decay amplitudes long after the decay occurs.  Otherwise,
for example, a $J/\psi\rightarrow e^+e^-$ decay at one nucleus
couldn't interfere with a $J/\psi\rightarrow$ hadrons at the other.
The observation of interference will be a clear demonstration that the
wave function doesn't collapse until after it is observed.  Here, the
'observation' occurs via an interaction with the beampipe surrounding
the collision region.

\section{Data Collection, Triggering and Analysis}\label{Expt}

We have studied exclusive $\rho^0\rightarrow\pi^+\pi^-$ production
with the STAR detector.  Data on gold-on-gold collisions at
$\sqrt{S_{NN}} = 130$\ GeV was collected during the Summer 2000 run.
In this data, collisions could occur every 210 nsec.  At STAR, the
luminosity reached a maximum of about $2\times10^{25}$/cm$^2$/s.

STAR detects charged particles in a 4.2 meter long time projection
chamber (TPC)\cite{TPC}.  The TPC has an inner radius of 50 cm, and an
outer radius of 2 m.  The TPC is centered around the interaction
region.  The pseudorapidity acceptance for charged particles depends
on the particle production point; for the data discussed here, the
interactions were spread longitudinally (in $z$), with $\sigma_z = 90$
cm.  The pseudorapidity ($y$) acceptance of the trigger and track
reconstruction depend on the individual vertex position. For a vertex
at $z=0$, the trigger was sensitive to charged particles with $|y|<1$;
the tracking covered a somewhat larger solid angle.  A solenoidal
magnet surrounds the TPC; for this data, the field was 0.25 T.  In
this field, the TPC momentum resolution was about $\Delta p/p = 2\% $.
Tracks with $p_T > 100$ MeV/c were reconstructed with good efficiency.
Tracks were identified by their energy loss in the TPC; this $dE/dx$
was measured with a resolution of about 8\%.

The TPC is surrounded by a cylindrical central trigger barrel (CTB).
For tracks from the center of the interaction region, it is sensitive
to tracks with $|y| < 1.0$.  This barrel consists of 240 scintillator
slats, each covering $\Delta y = 0.5$ by $\Delta\phi = \pi/30$.  The
scintillator light output is digitized to 8 bits of accuracy on each
crossing. In the 0.25 T magnetic field, the scintillator was
sensitive to charged particles with $p_T > 130$ MeV/c.  

Two zero degree calorimeters (ZDCs) at $z=\pm 18$ meters from the
interaction point detect neutrons from nuclear breakup\cite{zdc}.
These calorimeters are sensitive to single neutrons, and have an
acceptance of close to 100\% for neutrons from nuclear breakup.

The trigger hardware has several levels.  The initial Level 0 decision
uses lookup tables and field programmable gate arrays to initiate TPC
readout about 1.5 $\mu$s after the collision.  The other level used
here, Level 3, is based on on-line reconstruction using a small farm
of processors\cite{L3}; a level 3 acceptance triggered event building
and data transfer to tape.

\subsection{Triggering and Data Collection}

We studied $\rho^0$ production with two separate triggers.  The
topology trigger was designed to trigger on $\rho^0$ decay products
detected in the CTB system. A minimum bias trigger used the ZDCs to
select events where both nuclei dissociated.

The topology trigger was sensitive to a number of different
backgrounds.  The major ones were cosmic rays, beam gas interactions,
and debris from upstream interactions\cite{lund}.  The latter refers
to beam gas and/or beam-beampipe collisions far upstream from STAR;
the usual manifestation was one or more tracks in the TPC roughly
parallel to the beampipe, often accompanied by softer debris, perhaps
from interactions in the TPC wheel or magnetic pole tip.

\epsfclipon
\begin{figure}[htb]
\begin{center}
\leavevmode
\epsfysize= 5 cm
\epsfbox{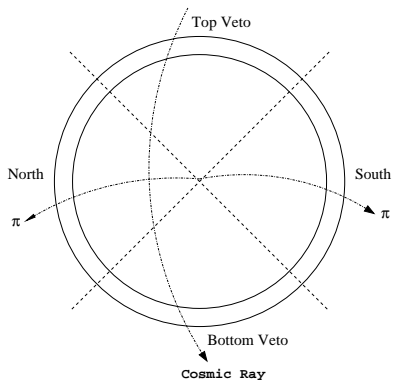}
\end{center}
\vskip -.2 in
\caption[]{Schematic view of the topology trigger.}
\label{fig:ptrigger}
\end{figure}

Fig. \ref{fig:ptrigger} shows how the topology trigger divided the CTB
into 4 azimuthal quadrants. It selected events with at least one hit
in the North and South sectors. The top and bottom quadrants were used
as vetos to reject most cosmic rays.  The trigger rate varied from 20
to 40 Hz, depending on luminosity.  Triggered events were scrutinized
by the Level 3 trigger, which reconstructed charged tracks online.
Events with more than 15 tracks, or with a vertex far outside the
interaction diamond were rejected.  The multiplicity cut removed
central collision events which left remnant energy in the trigger
detectors; this remnant energy could cause a topology trigger on
subsequent beam crossings.  The vertex cut removed debris from
upstream interactions, cosmic rays, and beam gas events.  These cuts
rejected 95\% of the events, leaving 1-2 events/second for event
building and storage.  In about 7 hours of data taking, we collected
about 30,000 events with this trigger.  The luminosity ranged from
about $2\times 10^{24}$/cm$^2$/s up to (briefly) $2\times
10^{25}$/cm$^2$/s,

The minimum bias trigger required a coincidence between the two zero
degree calorimeters.  The thresholds were set so that the efficiency
was high for single (or more) neutron deposition.  This trigger ran
throughout the STAR data taking.  This analysis is based on about
400,000 events.

\subsection{Exclusive $\rho^0$ Analysis}

Our analysis selected events with exactly two reconstructed tracks in
the TPC.  The tracks were vertexed with a low-multiplicity vertex
finder, and required to form a vertex within 2 cm of the TPC center in
$x$ and $y$, and within 2 m in $z$.  The vertexer projected the tracks
back from the TPC to the interaction region with allowance for
multiple scattering.  Events were accepted if the two tracks were
consistent with coming from a single vertex.

As the $|z|$ position of the vertex increases, the solid angle covered
by the TPC decreases, so events at larger $|z|$ are subject to more
background by higher-multiplicity processes where one or more tracks
is missed.  The major analysis backgrounds were grazing nuclear
collisions, incoherent photonuclear events, and beam gas events.  Some
cosmic rays also remained.

To reject the remaining cosmic rays, track pairs were required to be
at least slightly acoplanar, with a 3-dimensional opening angle less
than 3 radians.  This necessary cut reduced the $\rho^0$
reconstruction efficiency near $y=0$, where the two pion tracks are
nearly back-to-back.

Track $dE/dx$ was required to be consistent with that expected for a
pion.  Unfortunately, in the kinematic range for
$\rho\rightarrow\pi\pi$, the $\pi$ and $e$ $dE/dx$ bands overlap, and
only a few events were rejected.

\begin{figure}[htb]
\begin{center}
\leavevmode
\epsfysize=4.45 cm
\epsfbox{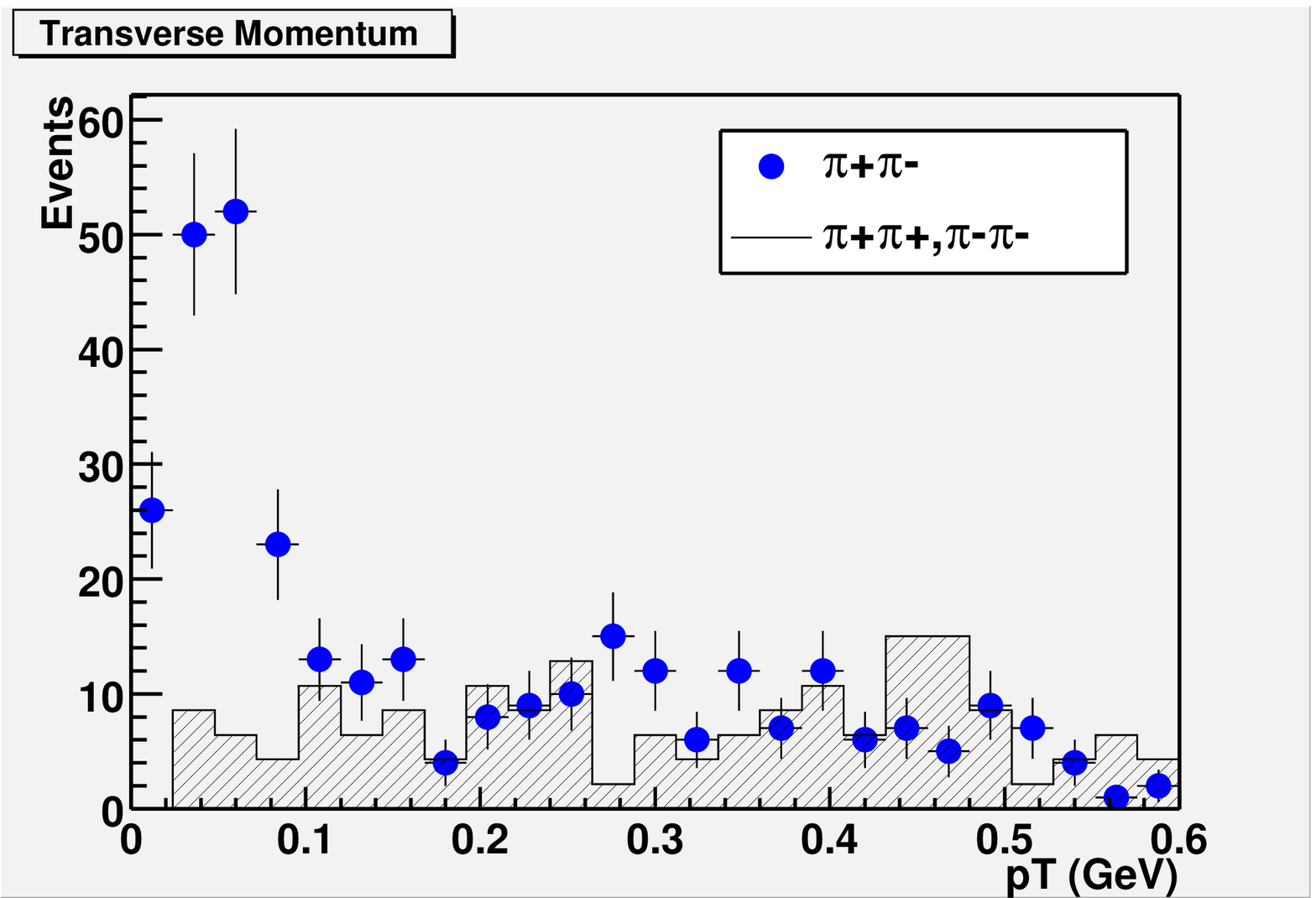}
\epsfysize=4.45 cm
\epsfbox{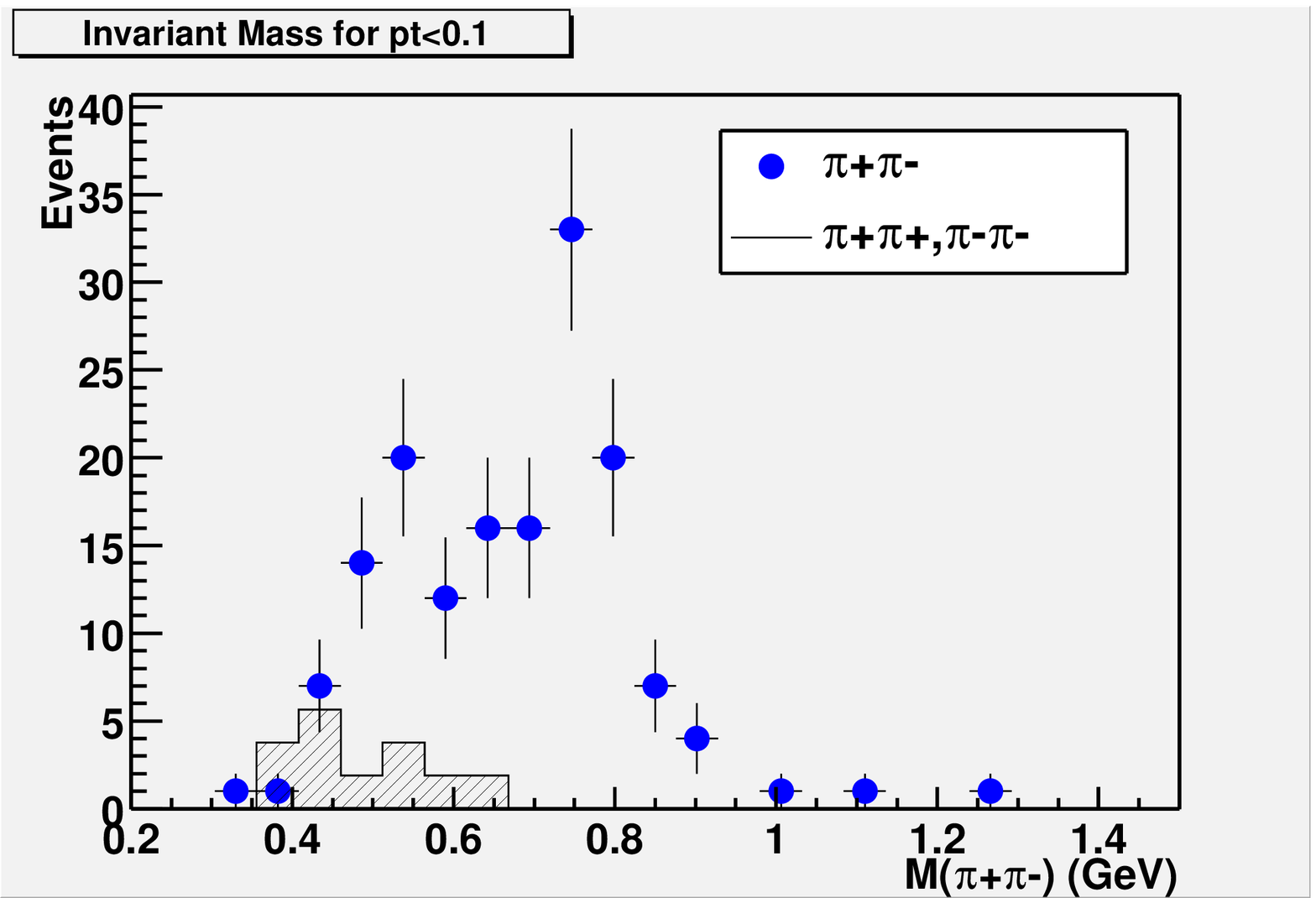}
\end{center}
\vskip -.1 in
\caption[]{(a) The $p_T$ spectrum of topology triggered 2-track
events.  (b) The $m_{\pi\pi}$ spectrum of 2-track events with $p_T <
100$ MeV/c.  The points are oppositely charged pairs, while the
histograms are the like-sign background, scaled up by a factor of
2.1.}
\vskip -.1 in
\label{fig:pctrig}
\end{figure}

Fig. \ref{fig:pctrig}(a) shows the $p_T$ spectrum of the topology
triggered pairs that pass these cuts, for unlike sign (dots - net
charge 0) and like sign pairs (histograms).  For the unlike pairs, a
large peak is visible at $p_T<100$ MeV/c.  This is consistent with
production that is coherent with both nuclei; the events with $p_T <
100$ MeV/c are considered our signal.  The like sign pairs have no
such enhancement, and can serve as a background sample.  The like-sign
pairs have been normalized to match the unlike sign in the signal-free
region 0.1 GeV $< m_{\pi\pi} < $ 1.0 GeV; this entailed scaling them
up by a factor of 2.1.

Fig. \ref{fig:pctrig}(b) shows the invariant mass of the pairs with
$p_T < 100$ MeV/c.  The points are the unlike sign events, while the
hatched histogram are the scaled like-sign pairs.  The like-sign pairs
are concentrated at relatively low masses, while the net charge 0
pairs have a peak around the rho mass.  We will consider the peak
shape later, for the summed $m_{\pi\pi}$ spectrum from both analyses.

\subsection{$\rho^0$ with Nuclear Excitation Analysis}

Events were selected by the minimum bias trigger, which took data
throughout the summer.  The acceptance for these events was
independent of the $\rho^0$ kinematics and the CTB calibration,
simplifying the interpretation.

\begin{figure}[htb]
\begin{center}
\leavevmode
\epsfysize=4.45 cm
\epsfbox{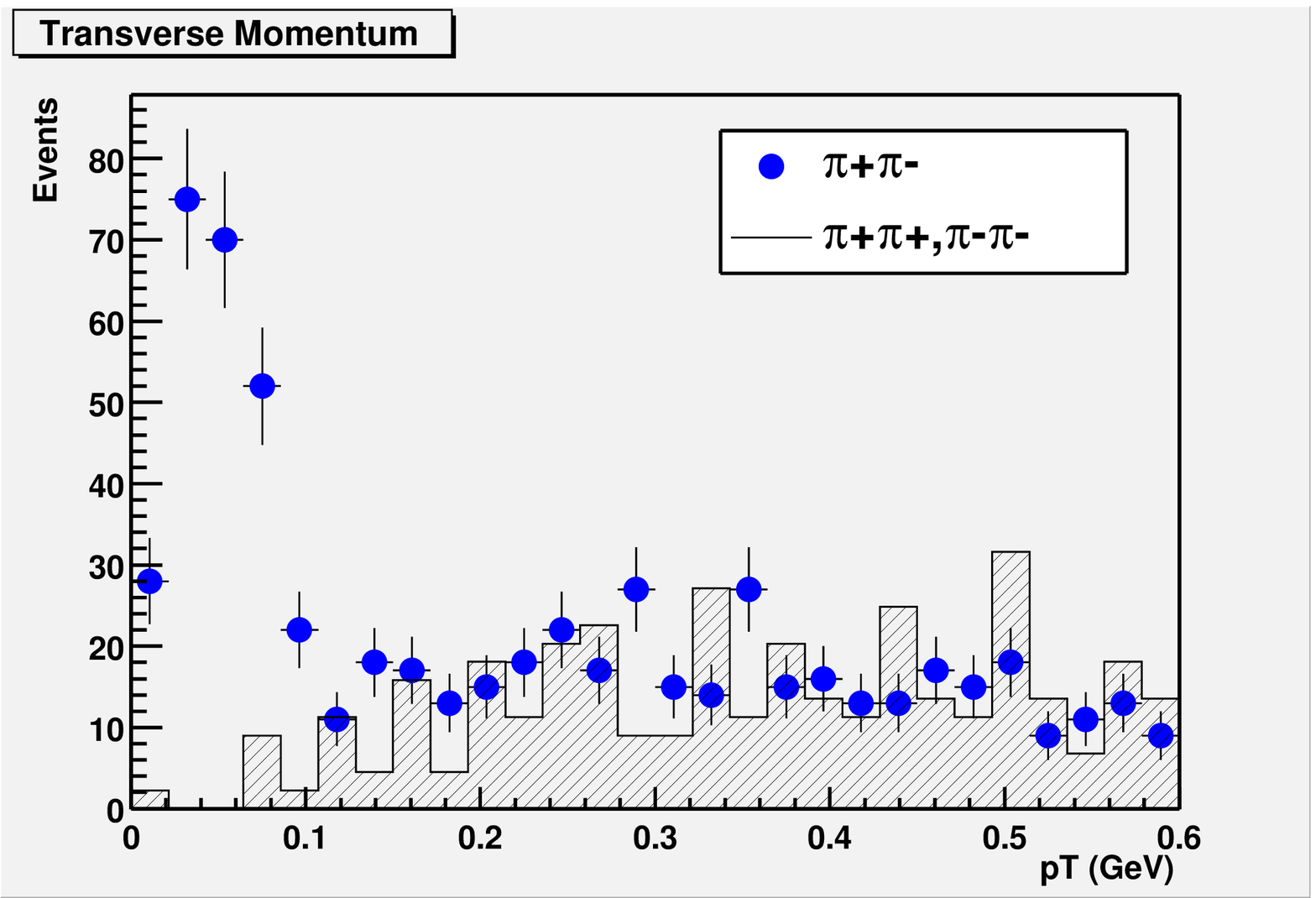}
\epsfysize=4.45 cm
\epsfbox{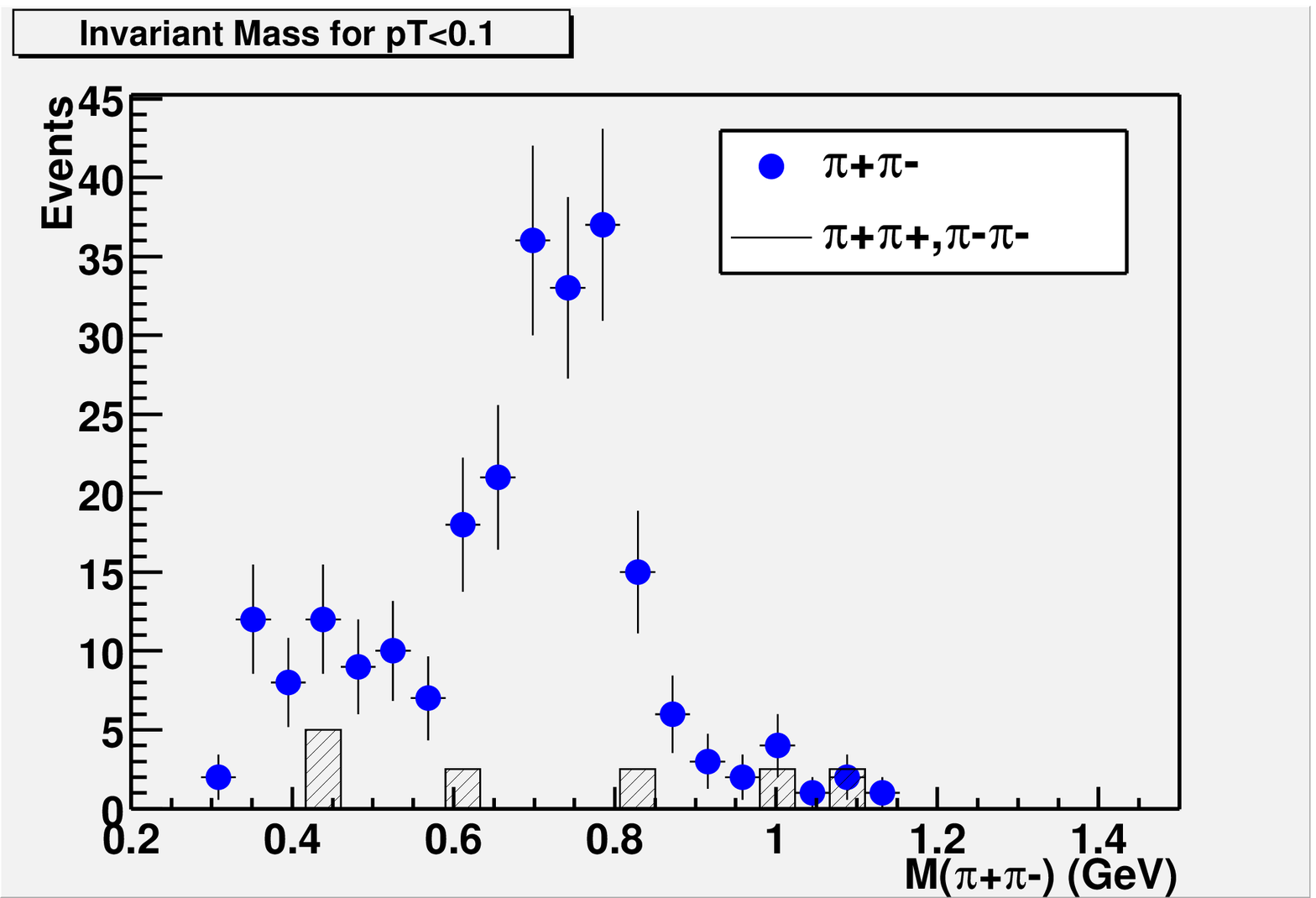}
\end{center}
\vskip -.1 in
\caption[]{(a) The $p_T$ spectrum of minimum bias 2-track events.  (b)
The $m_{\pi\pi}$ spectrum of 2-track events with $p_T < 100$ MeV/c.
The points are oppositely charged pairs, while the histograms are the
like-sign background, scaled up by 2.3.}
\vskip -.1 in
\label{fig:minbias}
\end{figure}

The event selection was the same as for the topology trigger sample.
Fig. \ref{fig:minbias}(a) shows the $p_T$ distribution of 2-track
events.  The opposite sign pairs show the same peak at $p_T< 100$
MeV/c as the topology triggered data. The like sign background
shows no peak, and, in fact, goes to zero as $p_T\rightarrow 0$, as
expected for a phase-space distribution.  The like-sign pairs were
scaled up to match the un-like sign pairs in the region 0.1 GeV $<
m_{\pi\pi} < $ 1.0 GeV; this scaling factor was 2.3.

This data contains little background from beam gas and incoherent
photonuclear events, which are unlikely to deposit energy in both
ZDCs.  However, the background from grazing nuclear collisions is
larger.

Fig. \ref{fig:minbias}(b) shows the $m_{\pi\pi}$ distribution for $p_T
< 100$ MeV/c in the same data, along with the scaled like-sign
background.  The signal is peaked around the $\rho$ mass.  

\begin{figure}[htb]
\begin{center}
\leavevmode
\epsfysize=5.5cm
\epsfbox{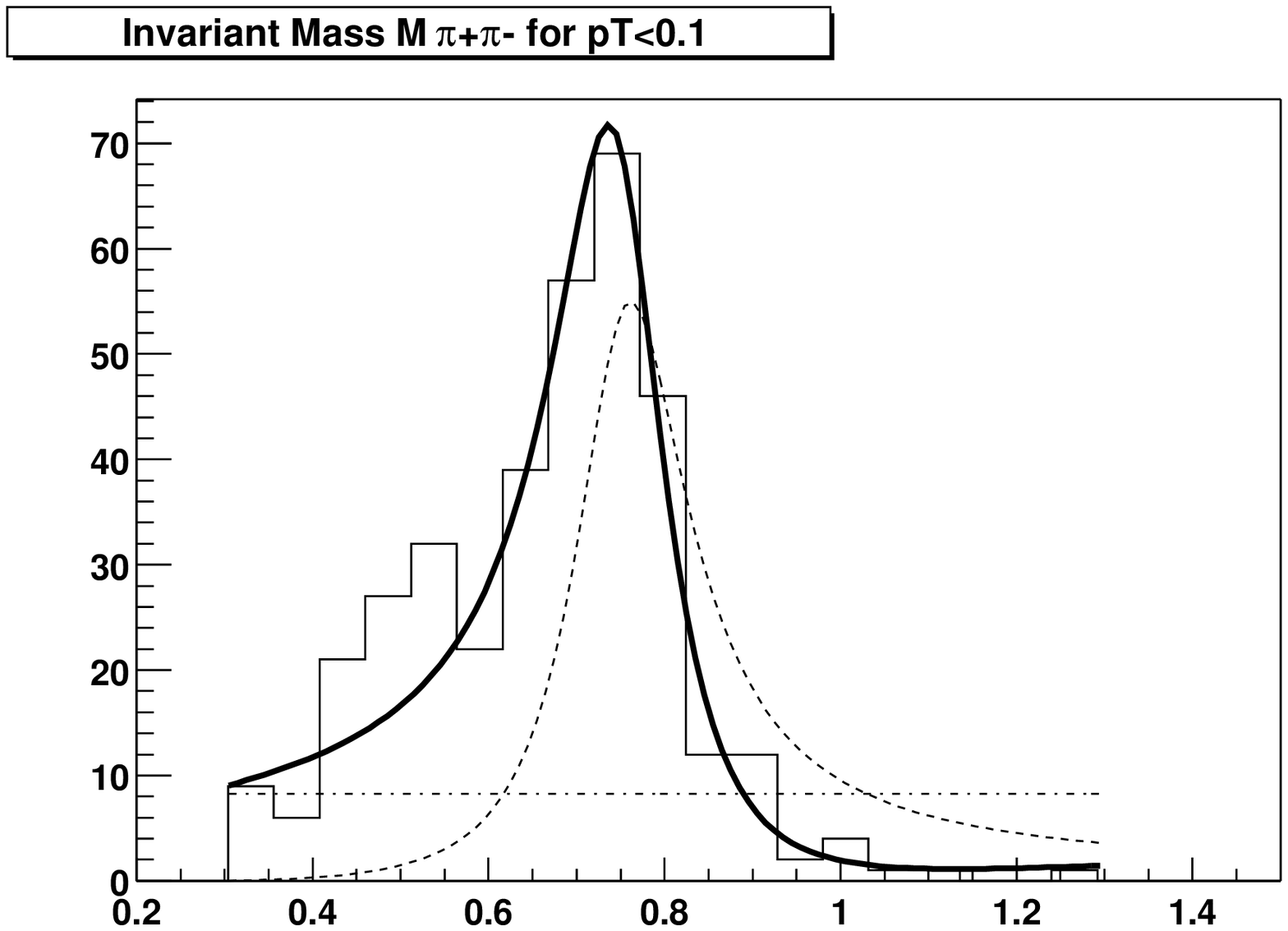}
\end{center}
\vskip -.15 in
\caption[]{The combined $m_{\pi\pi}$ invariant mass spectrum, with a fit
to $\rho$ and $\pi\pi$ components.  The histogram is the data.
The Breit-Wigner dashed curve is the $\rho$, the flat dashed
line direct $\pi^+\pi^-$, and the solid curve the combined fit
(including the interference).}
\vskip -.05 in
\label{fig:massspec}
\end{figure}

To improve the statistics, the minimum-bias and topology triggered
samples are combined into a single $m_{\pi\pi}$ spectrum, shown in
Fig. \ref{fig:massspec}.  The spectrum is fit to
Eq. (\ref{eq:rhosigma}).  In the fit, $M_\rho$ and $\Gamma_\rho$ are
fixed to minimize the number of free parameters.  However, if the mass
and width are allowed to float, the found values are consistent with
the actual values.  We find the relative size of the $\pi\pi$
contribution, $|B/A| = 0.89 \pm 0.07/\sqrt{\rm GeV}$ (statistical
error only).  This ratio is somewhat larger than that found by the
ZEUS collaboration, $|B/A| = 0.81 \pm 0.04 /\sqrt{\rm GeV}$ at the
same momentum transfer\cite{ZEUS}.  This is unexpected, since the
$\pi\pi$ component should have a higher nuclear absorption cross
section, so $|B/A|$ should drop as the target atomic number
increases\cite{ions}.

The background (not yet subtracted) may include a component from
electromagnetic $e^+e^-$ pair production, which would also be produced
with low $p_T$.  The background is concentrated at low $m_{\pi\pi}$,
so a subtraction will reduce the apparent $\pi\pi$ component.

\begin{figure}[htb]
\begin{center}
\leavevmode
\epsfysize=4.3 cm
\epsfbox{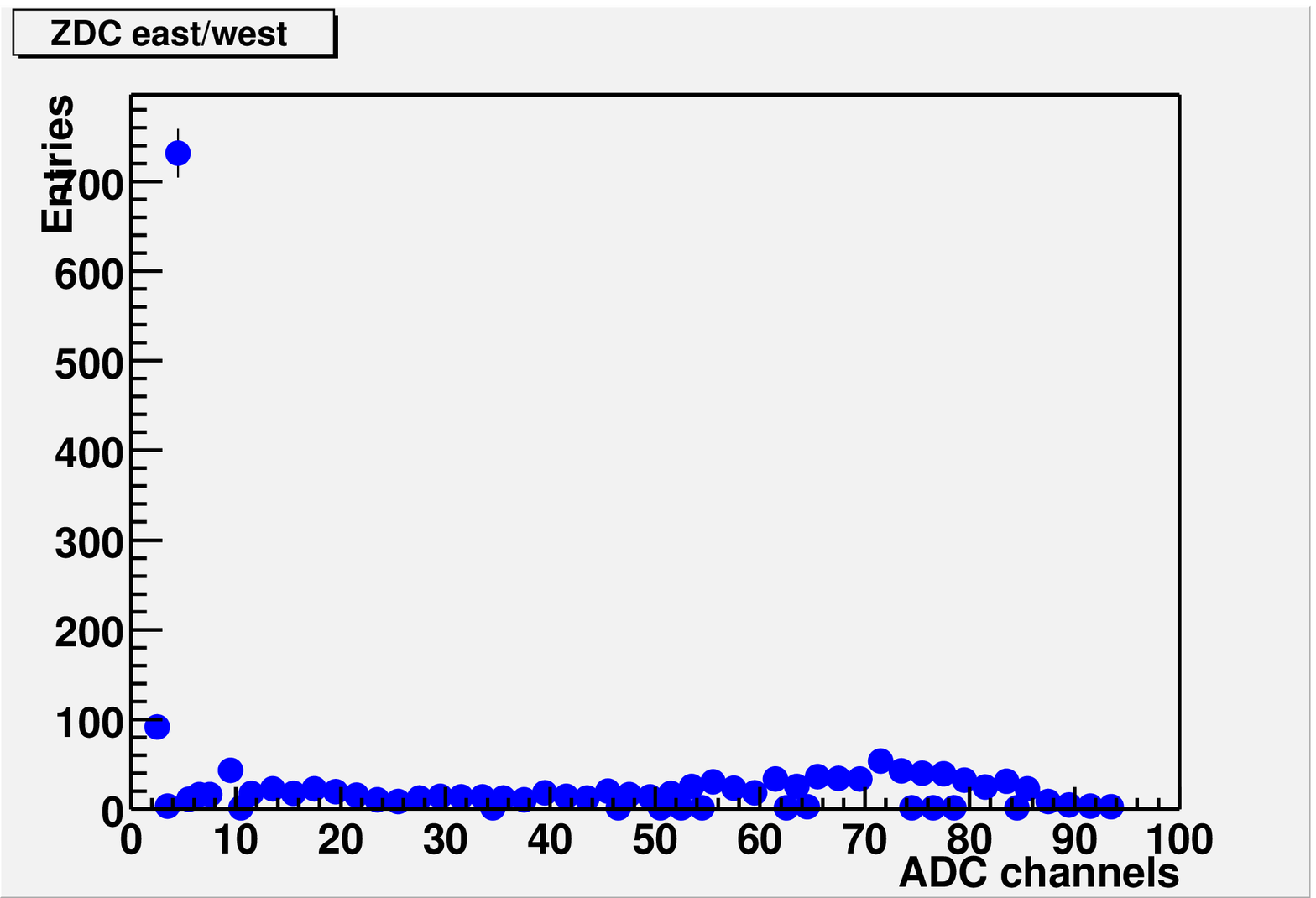}
\epsfysize=4.3 cm
\epsfbox{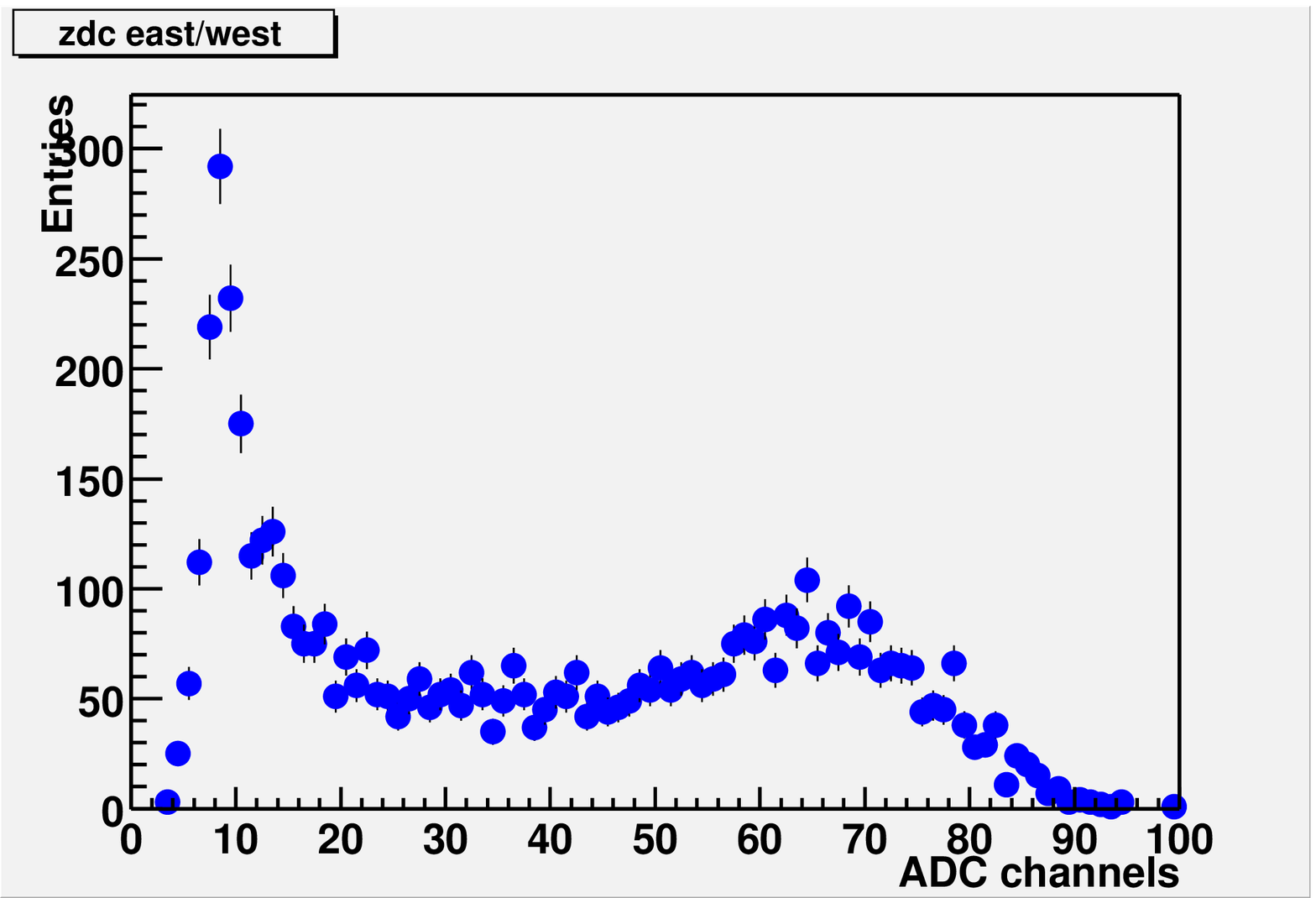}
\end{center}
\vskip -.2 in
\caption[]{(a) shows the ZDC energy distribution (arbitrary units) for
the topology triggered data.  The peak at 4 is the pedestal.  (b)
shows the ZDC energy distribution for minimum bias data.  The peak at
10 corresponds to single neutrons, while the higher-energy component
corresponds to multiple neutron emission.}
\label{fig:neutrons}
\vskip -.15 in
\end{figure}

Fig. \ref{fig:neutrons} compares the ZDC values from the two datasets.
The minimum bias data shows a clear single-neutron peak, with a
significant component at higher energies.  In contrast, the
topology triggered data shows almost no energy deposition.
This shows that the two processes are clearly distinguishable.

\section{Conclusions}

We have observed for the first time the reactions $Au +
Au \rightarrow Au + Au + \rho^0$ and $Au + Au \rightarrow Au^* + Au^*
+ \rho^0$.  The $\rho$ are produced at small perpendicular momentum,
showing their coherent coupling to both nuclei.

In the coming year, we expect to greatly expand the physics reach of
the STAR ultra-peripheral collisions program.  The improvement will
come from many factors, among them the increasing capability of the
STAR trigger, the additional detectors being installed in STAR, the
increased beam energy, and the increased luminosity and running time.
We plan to run the peripheral collisions trigger in parallel with the
central collisions trigger(s), greatly increasing our data collection
capabilities. The increased statistics will allow us to definitively
detector or rule out interference among the two vector meson
production sites.  We will also be able to study two-photon production
of mesons, and photoproduction of excited vector mesons.

\vskip -.15 in 
\section*{Notes} 
\begin{notes}
\vskip -.1 in
\item[a]
The collaboration membership is listed at
http://www.star.bnl.gov/STAR/smd\textunderscore l/collab/collab
\textunderscore sci\textunderscore apr01.ps.
\vskip -.1 in
\end{notes}

\vfill\eject

\vskip -.1 in
\begin{thebibliography}{99}  

\bibitem{rev1}G. Baur, K. Hencken and D. Trautmann, J. Phys. {\bf G
24}, 1657 (1998).

\bibitem{capri} S. Klein, preprint physics/0012021, to appear in
{\it Proc. 18th Advanced ICFA Beam Dynamics Workshop on Quantum
Aspects of Beam Physics}, World Scientific, 2001.

\bibitem{vmprod}S.R. Klein and J. Nystrand, Phys. Rev. C {\bf 60}, 014903
(1999).

\bibitem{soding}P. S\"oding, Phys. Lett. {\bf 19}, 702 (1966).

\bibitem{jackson}J. D. Jackson, Nuovo Cimento {\bf XXXV}, 6692 (1964).

\bibitem{pdg}D. E. Groom {\it et al.}, Eur. Phys. J, {\bf C15}, 1 (2000).

\bibitem{hencken}K. Hencken, D. Trautmann and G. Baur, Z. Phys {\bf
C68}, 473 (1995).

\bibitem{baltz}A. J. Baltz, M. J. Rhoades-Brown and J. Weneser,
Phys. Rev. {\bf E 54}, 4233 (1996).

\bibitem{vminter}S.R. Klein and J. Nystrand, Phys. Rev. Lett.
{\bf 84}, 2330 (2000).

\bibitem{TPC}H. Wieman {\it et al.}, IEEE Trans. Nucl.
Sci. {\bf 44}, 671 (1997).

\bibitem{zdc}C. Adler {\it et al.}, nucl-ex/0008005. 

%\bibitem{trigger}
%
\bibitem{L3}J. S. Lange {\it et al.}, Nucl. Instrum. Meth. {\bf A453},
397 (2000).

\bibitem{lund}J. Nystrand and S. Klein, nucl-ex/9811007, in {\it
Proc. Workshop on Photon Interactions and the Photon Structure}, eds.
G. Jarlskog and T. Sj\"ostrand, Lund, Sweden, September 1998.

\bibitem{ZEUS}J. Breitweg {\it et al.}, Eur. Phys. J {\bf C2},
247 (1998). The quoted $|B/A|$ ratio is from their Fig. 8a, at
$|t|=0$.

\bibitem{ions}G. McClellan {\it et al.}, Phys. Rev. {\bf D4}, 2683 (1971).

\end{thebibliography}
\end{document}